\documentclass[two-column, nofootinbib]{revtex4-1}

\usepackage{amsmath}
\usepackage{tabularx}
\usepackage{graphicx}
\usepackage[usenames,dvipsnames,svgnames]{xcolor}
\usepackage{hyperref}

\hypersetup{dvips,dvipdfm,colorlinks=true,urlcolor=magenta,filecolor=magenta,linktoc=page,citecolor=red,linkcolor=blue,bookmarks=true}

\begin{document}

\title{Thermodynamic implications of the gravitationally induced particle creation scenario}


\author{Subhajit Saha\footnote {Electronic Address: \texttt{\color{blue} subhajit1729@gmail.com}}}
\affiliation{Department of Physical Sciences, \\ Indian Institute of Science Education and Research Kolkata, Mohanpur 741246, West Bengal, India.} 

\author{Anindita Mondal\footnote{Electronic Address: \texttt{\color{blue}anindita12@bose.res.in}}}
\affiliation{Department of Astrophysics \& Cosmology, \\ S. N. Bose National Centre for Basic Sciences, Kolkata 700106, West Bengal, India.}


\begin{abstract}

\begin{center}
(Dated: The $27^{\text{th}}$ March, $2017$)
\end{center}

A rigorous thermodynamic analysis has been done at the apparent horizon of a spatially flat Friedmann-Lemaitre-Robertson-Walker universe for the gravitationally induced particle creation scenario with constant specific entropy and an arbitrary particle creation rate $\Gamma$. Assuming a perfect fluid equation of state $p=(\gamma -1)\rho$ with $\frac{2}{3} \leq \gamma \leq 2$, the first law, the generalized second law (GSL), and thermodynamic equilibrium have been studied and an expression for the total entropy (i.e., horizon entropy plus fluid entropy) has been obtained which does not contain $\Gamma$ explicitly. Moreover, a lower bound for the fluid temperature $T_f$ has also been found which is given by $T_f \geq 8\left(\frac{\frac{3\gamma}{2}-1}{\frac{2}{\gamma}-1}\right)H^2$. It has been shown that the GSL is satisfied for $\frac{\Gamma}{3H} \leq 1$. Further, when $\Gamma$ is constant, thermodynamic equilibrium is always possible for $\frac{1}{2}<\frac{\Gamma}{3H} < 1$, while for $\frac{\Gamma}{3H} \leq \text{min}\left\lbrace \frac{1}{2},\frac{2\gamma -2}{3\gamma -2} \right\rbrace$ and $\frac{\Gamma}{3H} \geq 1$, equilibrium can never be attained. Thermodynamic arguments also lead us to believe that during the radiation phase, $\Gamma \leq H$. When $\Gamma$ is not a constant, thermodynamic equilibrium holds if $\ddot{H} \geq \frac{27}{4}\gamma ^2 H^3 \left(1-\frac{\Gamma}{3H}\right)^2$, however, such a condition is by no means necessary for the attainment of equilibrium.

\end{abstract}

\maketitle



\section*{1. Introduction}

There have been several attempts to incorporate the present stage of cosmic acceleration into standard cosmology, the most notably being the introduction of an "exotic" matter termed dark energy (DE) which is believed to have a huge negative pressure, however, its nature and origin is still a mystery despite extensive research over the past one and a half decades. Several DE models have been proposed in the literature but observational data from various sources such as Supernovae Type Ia (SNe Ia), Cosmic Microwave Background (CMB), and Baryon Accoustic Oscillations (BAO) have established that the cosmological constant $\Lambda$ is the most viable candidate among them. The cosmic concordance $\Lambda$CDM model in which the Universe is believed to contain a cosmological constant $\Lambda$ associated with DE, and cold (i.e., pressureless) dark matter (abbreviated CDM) fits rather well the current astronomical data. 

Nevertheless, there are severe drawbacks corresponding to a finite but incredibly small value of $\Lambda$ such as the fine-tuning problem which leads to a discrepancy of $50$ to $120$ orders of magnitude with respect to its observed value which is about $3\times 10^{-11}\text{eV}^4$. Then there is the coincidence problem which is related to the question of "why are the energy densities of pressureless matter and DE of the same order precisely at the present epoch although they evolve so differently with expansion?" Several models such as decaying vacuum models, interacting scalar field descriptions of DE, and a single fluid model with an antifriction dynamics have been proposed with a view to alleviate such problems. Moreover, in order to solve the flatness and horizon problems, an inflationary stage for the very early universe was introduced but this again gave rise to several new problems, like the initial conditions, the graceful exit, and multiverse problems. 

Other attempts to explain the late time accelerating stage are modified gravity models, inhomogeneous cosmological models etc. but each one of them comes with several problems that are yet to be settled. Because of these said difficulties in various cosmological models, another well known proposal has been suggested --- the gravitationally induced particle creation mechanism. Schrodinger \cite{Schrodinger1} pioneered the microscopic description of such a mechanism which was further developed by Parker and others based on quantum field theory in curved spacetimes \cite{Parker00,Parker01,Birrell1,Mukhanov1,Parker1}. Prigogine and collaborators \cite{Prigogine1} provided a macroscopic description of particle creation mechanism induced by the gravitational field. A covariant description was later proposed \cite{Calvao1,Lima00} and the physical difference between particle creation and bulk viscosity was clarified \cite{Lima01}. The process of particle creation is classically described by introducing a backreaction term in the Einstein field equations whose negative pressure may provide a self-sustained mechanism of cosmic acceleration. Indeed, many phenomenological particle creation models have been proposed in the literature \cite{Zimdahl00,Gariel1,Abramo1,Lima02,Lima03,Alcaniz1}. It has also been shown that phenomenological particle production \cite{Lima04,Steigman1,Lima2,Basilakos1} can not only incorporate the late time cosmic acceleration but also provide a viable alternative to the concordance $\Lambda$CDM model. 

Despite rigorous investigation of various aspects of particle creation mechanism, its thermodynamic implications have never been explored. Such a study has been undertaken in this paper and the essence of this work is that the particle creation rate has been considered arbitrary, not a phenomenological one. The conclusions drawn from the present analysis are valid for any expression of the creation rate, constant or otherwise. The paper is organized as follows. Section 2 contains a brief discussion of the gravitationally induced adiabatic particle creation scenario, Section 3 along with Subsections A, B, and C are dedicated to detailed thermodynamic analysis of the process, while Section 4 provides a short discussion and possible scope for future work.

\section*{2. Gravitationally induced particle creation mechanism: A brief discussion} \label{sec2}

Let us consider a spatially flat, homogeneous and isotropic Friedmann-Lemaitre-Robertson-Walker (FLRW) universe with matter content endowed with the mechanism of particle creation. The dynamics of such a model is governed by the Friedman equations given by\footnote{In this manuscript, without any loss of generality, we have assumed that the physical constants, namely, $c$, $G$, $\hbar$, and $\kappa_B$, as well as $8\pi$ are unity.}
\begin{eqnarray}
3H^2 &=& \rho, \label{fr1} \\
\dot{H} &=& -\frac{1}{2}(\rho +p+\Pi). \label{fr2}
\end{eqnarray}
In the above equations, $\rho$ and $p$ are the energy density and thermostatic pressure of the cosmic fluid respectively and they are related by the equation of state (EoS) $p=(\gamma -1)\rho$ with $\frac{2}{3} \leq \gamma \leq 2$, $H=\frac{\dot{a}(t)}{a(t)}$ is the Hubble parameter [$a(t)$ is the scale factor of the Universe], and $\Pi$ is the creation pressure related to the gravitationally induced process of particle creation. The lower bound on $\gamma$ ensures that the perfect fluid does not become exotic, or equivalently, the strong energy condition remains valid. As a consequence, the energy conservation law gets reduced to
\begin{equation} \label{ece}
\dot{\rho}+\Theta (\rho +p+\Pi)=0.
\end{equation}

Now, the non-conservation of the total number $N$ of particles in an open thermodynamic system produces an equation given by
\begin{equation} \label{pce}
\dot{n}+\Theta n=n\Gamma.
\end{equation}
In Eqs. (\ref{ece}) and (\ref{pce}), $\Theta$ is the fluid expansion scalar which turns out to be $3H$ in our case, $\Gamma$ denotes the rate of change of the number of particles ($N=na^3$) in a comoving volume $a^3$, and $n$ is the number density of particles. So, a positive $\Gamma$ implies production of particles while a negative $\Gamma$ indicates particle annihilation. Further, a non-zero $\Gamma$ produces an effective bulk viscous pressure \cite{Hu1,Maartens1,Barrow1, Barrow2,Barrow3,Barrow4,Peacock1} of the fluid and hence non-equilibrium thermodynamics comes into the picture.

Using Eqs. (\ref{ece}) and (\ref{pce}), and the Gibb's relation
\begin{equation}
Tds=d\left(\frac{\rho}{n}\right)+pd\left(\frac{1}{n}\right),
\end{equation}
we can obtain an equation relating the creation pressure $\Pi$ and the creation rate $\Gamma$, which can be expressed as
\begin{equation}
\Pi =-\frac{\Gamma}{\Theta}(\rho +p),
\end{equation}
under the customary assumption that the specific entropy $s$ (in other words, the entropy per particle) is constant, i.e., the process is adiabatic (or isentropic). Thus, a dissipative fluid is equivalent to a perfect fluid with a non-conserved particle number. Eq.  (\ref{fr2}) now reduces to
\begin{equation} \label{dh/doth}
\frac{\dot{H}}{H^2}=-\frac{3\gamma}{2}\left(1-\frac{\Gamma}{3H}\right)
\end{equation}

The deceleration parameter $q$ takes the form
\begin{eqnarray}
q &=& -\frac{\dot{H}}{H^2}-1 \nonumber \\
&=& \frac{3\gamma}{2}\left(1-\frac{\Gamma}{3H}\right)-1,
\end{eqnarray} 
and the effective EoS parameter for this model (denoted by $w_{eff}$) becomes
\begin{eqnarray}
w_{eff} &=& \frac{p+\Pi}{\rho} \nonumber \\
&=& \gamma \left(1-\frac{\Gamma}{3H}\right)-1,
\end{eqnarray}
which represents quintessence era for $\Gamma <3H$ and phantom era for $\Gamma >3H$, while $\Gamma =3H$ corresponds to a cosmological constant, owing to the fact that $w_{eff}=-1$.

\section*{3. Thermodynamic analysis} \label{sec3}

In the following subsections, we shall study the first law, the generalized second law (GSL)\footnote{The idea of incorporating the GSL in cosmology was first developed by Ram Brustein \cite{Brustein1}. This second law is based on the conjecture that causal boundaries and not only event horizons have geometric entropies proportional to their area.}, and thermodynamic equilibrium for an arbitrary particle creation rate $\Gamma$. We shall consider an apparent horizon as our thermodynamic boundary, since, unlike the event horizon, a cosmic apparent horizon always exists and it coincides with the event horizon in the case of a last de Sitter expansion. Moreover, in a flat FLRW universe, the apparent horizon coincides with the Hubble horizon $H^{-1}$. So, the apparent horizon can be considered to be located at $R_{A}=\frac{1}{H}$ and its first order derivative with respect to the cosmic time $t$ can be evaluated 
as
\begin{eqnarray} \label{dra}
\dot{R}_A &=& -\frac{\dot{H}}{H^2} \nonumber \\
&=& \frac{3\gamma}{2}\left(1-\frac{\Gamma}{3H}\right).
\end{eqnarray}
The (Bekenstein) entropy and (Hawking) temperature of the apparent horizon are given by
\begin{eqnarray} \label{sa}
S_A &=& \left(\frac{c^3}{G\hbar}\right)\frac{4\pi R_{A}^{2}}{4} \nonumber \\
&=& \frac{1}{8}R_{A}^{2},
\end{eqnarray}
and
\begin{eqnarray} \label{ta}
T_A &=& \left(\frac{\hbar c}{\kappa _B}\right)\frac{1}{2\pi R_A} \nonumber \\
&=& \frac{4}{R_A}
\end{eqnarray}
respectively.

\subsection{First law} \label{subseca}

The first law of thermodynamics at the horizon is governed by the Clausius relation
\begin{equation}
-dE_A=T_AdS_A.
\end{equation} 
The differential $dE_A$ of the amount of energy crossing the apparent horizon can be evaluated as (see Eq. (27) of Ref. \cite{Bousso1})
\begin{eqnarray} \label{dea}
-dE_A &=& \frac{1}{2}R_{A}^{3}(\rho +p)Hdt \nonumber \\
&=& \frac{3\gamma}{2}dt.
\end{eqnarray}

Again, using the expressions of $T_A$ and $S_A$ given in Eqs. (\ref{sa}) and (\ref{ta}), the expression $T_AdS_A$ becomes
\begin{equation}
T_AdS_A=\frac{3\gamma}{2}\left(1-\frac{\Gamma}{3H}\right)dt,
\end{equation}
where we have used relation (\ref{dra}).

From the above analysis, we find that the first law holds at the apparent horizon whenever $\Gamma =0$, or loosely speaking, whenever $\Gamma \ll 3H$.

\subsection{Generalized second law: An expression for total entropy} \label{subsecb}

According to thermodynamics, the equilibrium configuration of an isolated macroscopic physical system should
be the maximum entropy state, consistent with the constraints imposed on the system. Thus if $S$ is the total entropy of the system, the following conditions should hold --- (a) $dS \geq 0$ [i.e., the entropy function cannot decrease (the second law of thermodynamics)], and (b) $d^2 S < 0$ [i.e., the entropy function attains a maximum (thermodynamic equilibrium)]. In our context, the Universe bounded by an apparent horizon and filled with some cosmic fluid forms an isolated macroscopic physical system for which the above inequalities can be generalized as
\begin{equation}
\text{(i)}~d(S_A+S_f) \geq 0 ~~~~~~~~\text{and}~~~~~~~~ \text{(ii)}~d^2 (S_A+S_f) < 0
\end{equation}
respectively, where $S_f$ is the entropy of the cosmic fluid contained within the horizon. The inequality (i) is sometimes called the GSL.

The Gibb's equation can be rewritten in the form
\begin{equation} \label{geq}
T_fdS_f=dE_f+pdV,
\end{equation}
where $T_f$ is the temperature of the cosmic fluid respectively, and $E_f=\rho V$ is the energy of the fluid. 

Now, the assumption of a constant specific entropy leads us to an evolution equation for the fluid temperature given by (see the second relation in Eq. (35) of Ref. \cite{Zimdahl1})
\begin{equation}
\frac{\dot{T}_f}{T_f}=(\Gamma - \Theta)\frac{\partial p}{\partial \rho}.
\end{equation}
Noting from Eq. (\ref{dh/doth}) that $(\Gamma - \Theta) = \frac{2}{\gamma}\left(\frac{\dot{H}}{H}\right)$, the above equation leads to the integral
\begin{equation}
\text{ln}\left(\frac{T_f}{T_0}\right)=\frac{2 (\gamma -1)}{\gamma} \int \frac{dH}{H}. \nonumber
\end{equation}
On integration, we obtain,
\begin{equation}
T_f=T_0 H^{\frac{2(\gamma -1)}{\gamma}},
\end{equation}
where $T_0$ is the constant of integration. Note that $\Gamma$ does not appear explicitly in the equation.

From Eq. (\ref{geq}), the differential of the fluid entropy can be obtained in the following form:
\begin{equation}
dS_f=\frac{3\gamma}{2}\left(\frac{3\gamma}{2}-1\right)T_{0}^{-1}\left(1-\frac{\Gamma}{3H}\right)H^{\frac{2(1-\gamma)}{\gamma}}dt.
\end{equation}
The differential of the total entropy can then be evaluated as
\begin{equation} \label{st}
d(S_A+S_f)=\frac{3\gamma}{8H}\left(1-\frac{\Gamma}{3H}\right)\left[1+4\left(\frac{3\gamma}{2}-1\right)T_{0}^{-1}H^{\frac{2}{\gamma}-1}\right]dt.
\end{equation}
It can be easily seen from the previous equation that GSL holds if $\Gamma \leq 3H$, or equivalently, if $\frac{\Gamma}{3H} \leq 1$. Therefore, the GSL is not consistent with the phantom fluid. Furthermore, $S_T$ is a constant of motion when $\Gamma =3H$, i.e., when $w_{eff}=-1$, a cosmological constant.

\begin{figure}
\includegraphics[height=3in,width=3in]{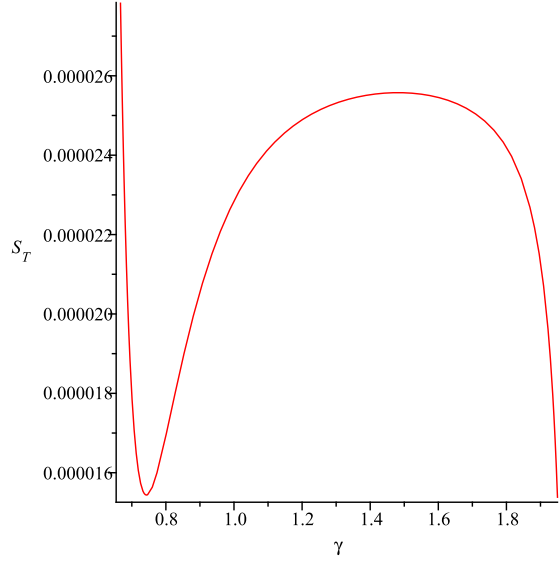}
\caption{The total entropy $S_T$ for allowed values of $\gamma$, taking $H=67$ and $T_0=10^{5}$.} \label{fig01}
\end{figure}

Another remarkable fact is that Eq. (\ref{st}) gives us an opportunity [by replacing $dt$ by $\frac{dH}{\dot{H}}$ and using Eq. (\ref{dh/doth})] to derive an expression for the total entropy in terms of the Hubble parameter $H$ in the form
\begin{eqnarray}
S_T &=& S_A+S_f \nonumber \\
&=& \frac{1}{8H^2}\left[1-8\left(\frac{\frac{3\gamma}{2}-1}{\frac{2}{\gamma}-1}\right)T_{0}^{-1}H^{\frac{2}{\gamma}}\right]. \label{sth}
\end{eqnarray}
The essence of Eq. (\ref{sth}) lies in the fact that the particle creation rate $\Gamma$ does not occur explicitly in the equation. Requiring that the total entropy be always positive, we can, in principle, obtain a lower bound on $T_{0}$ given by
\begin{equation} \label{to}
T_0 \geq 8\left(\frac{\frac{3\gamma}{2}-1}{\frac{2}{\gamma}-1}\right)H^{\frac{2}{\gamma}}.
\end{equation}
Eq. (\ref{to}) implies that we can also impose a lower bound on the fluid temperature $T_f$ as
\begin{equation} \label{tflb}
T_f \geq 8\left(\frac{\frac{3\gamma}{2}-1}{\frac{2}{\gamma}-1}\right)H^2.
\end{equation}
For radiation era (i.e., $\gamma =\frac{4}{3}$) and matter dominated era (i.e., $\gamma =1$), the lower bounds on $T_f$ become $T_f \geq 16H^2$ and $T_f \geq 4H^2$ respectively. Using Maple software, the total entropy $S_T$ has been plotted against $\gamma$ for $H=67$ and $T_0=10^{5}$, and presented in Figure \ref{fig01}. 

\subsection{Thermodynamic equilibrium} \label{subsecc}

{\bf Case I:} $\Gamma$ is constant --- If the particle creation rate $\Gamma$ is assumed to be constant, then the second order differential of the total entropy can be obtained from Eq. (\ref{st}) as
\begin{eqnarray} \label{d2st}
\frac{d^2S_T}{dt^2} &=& \frac{d^2}{dt^2}(S_A+S_f) \nonumber \\
&=& \frac{9\gamma ^2}{16}\left(1-\frac{\Gamma}{3H}\right)\left(1-\frac{2\Gamma}{3H}\right)\left[1+4\left(\frac{3\gamma}{2}-1\right) T_{0}^{-1}H^{\frac{2}{\gamma}-1}\left\lbrace 1-\left(\frac{2}{\gamma}-1\right)\left(\frac{1-\frac{\Gamma}{3H}}{1-\frac{2\Gamma}{3H}}\right)\right\rbrace \right].
\end{eqnarray}
In Table I, we have explored relevant subintervals of $\Gamma$ in order to test for the validity of thermodynamic equilibrium. From the table, it is evident that thermodynamic equilibrium holds unconditionally for $\frac{1}{2}<\frac{\Gamma}{3H}<1$, while it never holds for $\frac{\Gamma}{3H} \leq \text{min}\left\lbrace \frac{1}{2},\frac{2\gamma -2}{3\gamma -2} \right\rbrace$ and $\frac{\Gamma}{3H} \geq 1$. Thus, thermodynamic equilibrium in this case is inconsistent with the cosmological constant as well as the phantom fluid.  

From different observational sources, it has been well established that the radiation phase was followed by a matter dominated era which eventually transited to a second de Sitter phase. Accordingly, it can be expected that in the radiation dominated era, the entropy increased but the thermodynamic equilibrium was not achieved \cite{Lima1}. If this were not true, the Universe would have attained a state of maximum entropy and would have stayed in it forever unless acted upon by some "external agent." However, it is a well known fact \cite{Parker1} that the production of particles was suppressed during the radiation phase, so in this model, there would be no external agent to remove the system from thermodynamic equilibrium. Therefore, our present analysis leads us to conclude that during the radiation phase, if $\Gamma$ is constant, then $\frac{\Gamma}{3H} \leq \frac{1}{3}$, or equivalently, $\Gamma \leq H$.\\
\begin{center} 
{\bf Table I}: Equilibrium configuration for different subintervals of $\Gamma$ 
\end{center}
\begin{center}
\begin{tabular}{|c|c|c|c|}
\hline Subintervals of $\Gamma$ & Sign of $\left(1-\frac{\Gamma}{3H}\right)\left(1-\frac{2\Gamma}{3H}\right)$ & Sign of $\left\lbrace 1-\left(\frac{2}{\gamma}-1\right)\left(\frac{1-\frac{\Gamma}{3H}}{1-\frac{2\Gamma}{3H}}\right) \right\rbrace$ & Equilibrium? \\
\hline \hline $\Gamma \leq \frac{3H}{2}$ & Non-negative & Non-negative for $\frac{\Gamma}{3H} < \frac{2\gamma -2}{3\gamma -2}$ & Never for $\frac{\Gamma}{3H} \leq \text{min}\left\lbrace \frac{1}{2},\frac{2\gamma -2}{3\gamma -2} \right\rbrace$\\
\hline $\frac{3H}{2}<\Gamma<3H$ & Negative & Positive & Always \\
\hline $\Gamma \geq 3H$ & Non-Negative & Positive & Never \\
\hline
\end{tabular}
\end{center}

\vspace*{0.4cm}

{\bf Case II:} $\Gamma$ is not constant --- For a variable $\Gamma$, Eq. (\ref{d2st}) can be generalized as
\begin{eqnarray} \label{d2stg}
\frac{d^2S_T}{dt^2} &=& \frac{27\gamma ^2}{16} \Biggl[\left\lbrace \left(1-\frac{\Gamma}{3H}\right)^2-\left(\frac{4}{27\gamma ^2}\right)\frac{\ddot{H}}{H^3}\right\rbrace \left\lbrace 1+4\left(\frac{3\gamma}{2}-1\right)T_{0}^{-1}H^{\frac{2}{\gamma}-1}\right\rbrace -\frac{4}{3}\left(\frac{3\gamma}{2}-1\right)\left(\frac{2}{\gamma}-1\right) \nonumber \\
&\times& T_{0}^{-1}H^{\frac{2}{\gamma}-1}\left(1-\frac{\Gamma}{3H}\right)^2 \Biggr],
\end{eqnarray}
where we have substituted the value of $\dot{\Gamma}$ evaluated as
$$\dot{\Gamma}=(6H-\Gamma)\frac{\dot{H}}{H}+\left(\frac{2}{\gamma}\right)\frac{\ddot{H}}{H}.$$
It is evident from Eq. (\ref{d2stg}) that it is quite difficult to perform an analysis similar to the one that we have done in the previous case. The only definite conclusion which can be made here is that the thermodynamic equilibrium holds if $\ddot{H} \geq \frac{27}{4}\gamma ^2 H^3 \left(1-\frac{\Gamma}{3H}\right)^2$.

\section*{4. Discussion and future work} \label{sec4}

This paper dealt with a rigorous thermodynamic analysis at the apparent horizon of a spatially flat FLRW universe for the gravitationally induced particle creation scenario with constant specific entropy and an arbitrary particle creation rate $\Gamma$. Assuming a perfect fluid EoS $p=(\gamma -1)\rho$ with $\frac{2}{3} \leq \gamma \leq 2$ (the lower bound ensures that the strong energy condition remains valid), the first law, the GSL, and thermodynamic equilibrium have been studied and the following results have been found---
\begin{itemize}
\item The first law holds at the apparent horizon either for a zero particle creation rate or, loosely speaking, when the creation rate is infinitesimally small as compared to $3H$.
\item The GSL holds if $\Gamma \leq 3H$, or equivalently, if $\frac{\Gamma}{3H} \leq 1$, which implies that the GSL is not consistent with the phantom fluid.
\item For a constant particle creation rate, thermodynamic equilibrium always holds for $\frac{1}{2}<\frac{\Gamma}{3H}<1$, while it never holds for $\frac{\Gamma}{3H} \leq \text{min}\left\lbrace \frac{1}{2},\frac{2\gamma -2}{3\gamma -2} \right\rbrace$ and $\frac{\Gamma}{3H} \geq 1$. Thus, thermodynamic equilibrium in this case is inconsistent with the cosmological constant as well as the phantom fluid.
\item When $\Gamma$ is not constant, the only definite conclusion which can be made is that the thermodynamic equilibrium holds if $\ddot{H} \geq \frac{27}{4}\gamma ^2 H^3 \left(1-\frac{\Gamma}{3H}\right)^2$, however, such a condition is by no means necessary for the attainment of equilibrium.
\end{itemize}

An expression for the total entropy with no explicit dependence on $\Gamma$ has also been found. Such an expression suggests that for $\Gamma =3H$, i.e., a cosmological constant, the total entropy is a constant of motion. Further, imposing the condition that the total entropy is always positive, a lower bound on the fluid temperature $T_f$ has been obtained. It is evident that $T_f \geq 16H^2$ and $T_f \geq 4H^2$ for radiation and matter dominated eras respectively. Thermodynamic arguments also lead us to believe that if $\Gamma$ is a constant, then $\Gamma \leq H$ during the radiation phase.

For future work, thermodynamics of the particle creation scenario at any arbitrary horizon can be investigated. The present thermodynamic analysis can also help to constrain various parameters of phenomenological particle creation rates that have been considered in recent literature \cite{Lima1,Steigman1,Lima2,Lima3,Fabris1,Chakraborty1,Chakraborty00,Saha1,Chakraborty2,Nunes1}. Further, attempts to incorporate matter creation in inhomogeneous cosmological models can be made and its thermodynamic implications can be studied.


\begin{acknowledgments}

Subhajit Saha is supported by SERB, Govt.of India under National Post-doctoral Fellowship Scheme [File No. PDF/2015/000906]. Anindita Mondal wishes to thank DST, Govt. of India for providing Senior Research Fellowship. The authors are thankful to the anonymous reviewer for insightful comments which have helped to improve the quality of the manuscript significantly.

\end{acknowledgments}


\end{document}